\documentclass[journal]{IEEEtran}

\usepackage{cite}
\usepackage{hyperref}
\usepackage{amsmath,amssymb,amsfonts}
\usepackage{algorithmic}
\usepackage{graphicx}
\usepackage{subfigure}
\usepackage{textcomp}
\usepackage{xcolor}
\usepackage{multirow}

\makeatletter
\renewcommand{\@thesubfigure}{\hskip\subfiglabelskip}
\makeatother

\begin{document}

\title{Joint Reference Frame Synthesis and Post Filter Enhancement for Versatile Video Coding}

\author{
Weijie~Bao,
Yuantong~Zhang,
Jianghao~Jia,
Zhenzhong~Chen,~\IEEEmembership{Senior~Member,~IEEE},
Shan~Liu,~\IEEEmembership{Fellow,~IEEE}

\thanks{
(Corresponding author: Zhenzhong Chen)
}
}

\maketitle

\begin{abstract}
This paper presents the joint reference frame synthesis (RFS) and post-processing filter enhancement (PFE) for Versatile Video Coding (VVC), aiming to explore the combination of different neural network-based video coding (NNVC) tools to better utilize the hierarchical bi-directional coding structure of VVC. Both RFS and PFE utilize the Space-Time Enhancement Network (STENet), which receives two input frames with artifacts and produces two enhanced frames with suppressed artifacts, along with an intermediate synthesized frame. STENet comprises two pipelines, the synthesis pipeline and the enhancement pipeline, tailored for different purposes. During RFS, two reconstructed frames are sent into STENet's synthesis pipeline to synthesize a virtual reference frame, similar to the current to-be-coded frame. The synthesized frame serves as an additional reference frame inserted into the reference picture list (RPL). During PFE, two reconstructed frames are fed into STENet's enhancement pipeline to alleviate their artifacts and distortions, resulting in enhanced frames with reduced artifacts and distortions. To reduce inference complexity, we propose joint inference of RFS and PFE (JISE), achieved through a single execution of STENet. Integrated into the VVC reference software VTM-15.0, RFS, PFE, and JISE are coordinated within a novel Space-Time Enhancement Window (STEW) under Random Access (RA) configuration. The proposed method could achieve -7.34\%/-17.21\%/-16.65\% PSNR-based BD-rate on average for three components under RA configuration.
\end{abstract}


\IEEEpeerreviewmaketitle

\section{Introduction}

\IEEEPARstart{O}{ver} the past few decades, numerous video coding standards have emerged to compress the ever-expanding volume of video data. The latest video coding standard Versatile Video Coding (VVC) was released in July 2020 by the Joint Video Experts Team (JVET), which was founded by two international standardization organizations, ITU-T Video Coding Experts Group (VCEG) and ISO/IEC Moving Picture Experts Group (MPEG)\cite{Bross-TCSVT2021}. Compared to the previous video standard High Efficiency Video Coding (HEVC), VVC reaches approximately 50\% bit rate reduction for nearly equal video quality\cite{Bross-Proceedings2021}. Moreover, VVC is more versatile for certain types of video content, such as ultra-high resolution videos, 360° videos, and gaming content videos\cite{Bross-TCSVT2021}.

With the development of deep learning, neural network-based video coding (NNVC) has made tremendous progress. The existing works can be classified into two categories: deep schemes and deep tools\cite{Liu-ACM2021}. The deep schemes are built primarily upon end-to-end deep networks as encoders and decoders. Deep schemes have garnered significant attention due to their promising prospects\cite{Guo-CVPR2019, Li-CVPR2023, Li-CVPR2024}. On the other hand, the deep tools (NNVC tools) shall be used within traditional coding schemes or together with traditional coding tools. These NNVC tools can provide additional performance enhancements when integrated into traditional coding schemes.

Inter prediction is a crucial part of block-based hybrid coding schemes that reduce temporal redundancy in video sequences. During inter prediction, the to-be-coded frame will select frames from reference picture lists (RPLs) as reference frames for motion estimation (ME) and motion compensation (MC)\cite{Chien-TCSVT2021}. Reference frames much more similar to the to-be-coded frame will lead to smaller residuals in the bitstream. This characteristic inspires some works to enhance inter prediction by synthesizing reference frames\cite{Zhao-ICIP2018,Guo-VCIP2020,Hu-ACM2023,Jia-TCSVT2023}. During inter prediction, different kinds of reference frame synthesis networks are required for bi-directional prediction and uni-directional prediction. Specifically, interpolation networks enhance bi-directional prediction, while extrapolation networks improve uni-directional prediction.

Meanwhile, block partitioning and quantization in block-based hybrid coding schemes lead to artifacts in reconstructed frames, including block artifacts, ringing artifacts, and over-smoothing. In VVC, various in-loop filters such as deblocking filter (DBF), sample adaptive offset (SAO), adaptive loop filtering (ALF), cross-component adaptive loop filtering (CC-ALF), and luma mapping with chroma scaling (LMCS) are employed to mitigate these artifacts\cite{Karczewicz-TCSVT2021}. Nowadays, some work has demonstrated that combining neural network-based in-loop filters with handcrafted in-loop filters can lead to significant coding improvements\cite{Yue-ICIP2021,Wang-CVPRW2022}. Meanwhile, some researchers have introduced neural network-based post-processing filters designed to improve video quality\cite{Zhang-ICME2020,Ma-SP2020,Santamaria-ISM2021,Zhang-IEEEAccess2023}. Different from neural network-based in-loop filters, neural network-based post-processing filters can apply to any video standard without any codec modifications.

\begin{figure*}
\centerline{\includegraphics[width=18cm]{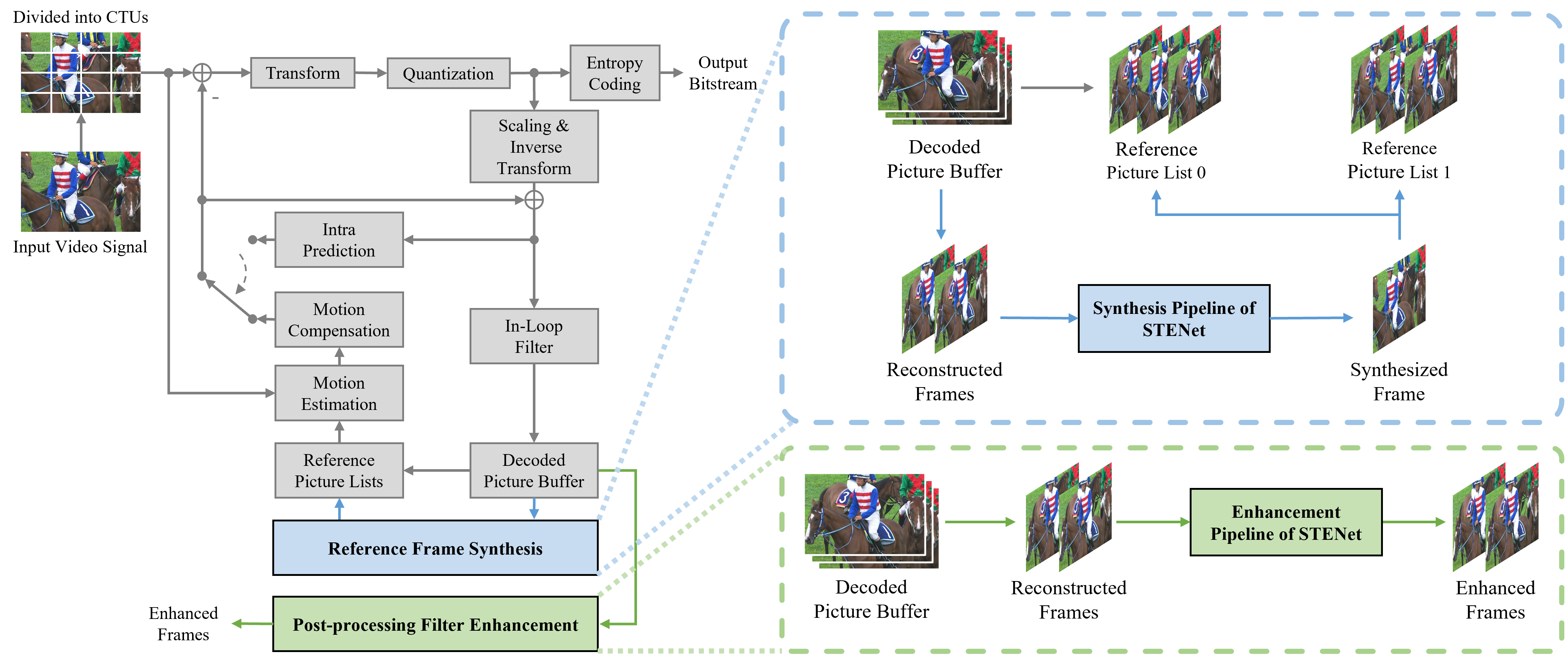}}
\caption{The framework of joint reference frame synthesis (RFS) and post-processing filter enhancement (PFE). During RFS, two reconstructed frames from DPB are input into STENet's synthesis pipeline to synthesize an intermediate frame, treated as the virtual reference frame, and inserted into two RPLs. During PFE, two reconstructed frames are selected from DPB and input into STENet's enhancement pipeline to alleviate their artifacts and distortions, resulting in higher-quality final output frames.}
\label{Framework}
\end{figure*}

In this paper, we propose a novel approach: joint reference frame synthesis (RFS) and post-processing filter enhancement (PFE) for VVC, aiming to explore the combination of different NNVC tools to better utilize the hierarchical bi-directional coding structure of VVC. By combining RFS and PFE, the proposed method achieves space-time coupled enhancement among neighboring frames. In VVC, Random Access (RA) configuration defines a hierarchical bi-directional coding structure, which effectively utilizes temporal information in contrast to both Low Delay P (LDP) and Low Delay B (LDB) configurations. Considering this, the paper focuses on RA configuration to better utilize bi-directional temporal information for space-time coupled enhancement. The overall framework is shown in Fig.\ref{Framework}. Both RFS and PFE utilize the Space-Time Enhancement Network (STENet), illustrated in Fig.\ref{Network Architecture}. STENet takes two compressed frames with artifacts and distortions, fully leveraging the space-time information to generate an intermediate synthesized frame and two enhanced frames with reduced artifacts. STENet comprises two pipelines: the synthesis pipeline for RFS and the enhancement pipeline for PFE. During RFS, two reconstructed frames from the decoded picture buffer (DPB) are input into STENet's synthesis pipeline to synthesize an intermediate frame, treated as the virtual reference frame, and inserted into two RPLs. Meanwhile, during PFE, two reconstructed frames are selected from DPB and input into STENet's enhancement pipeline to alleviate their artifacts and distortions, resulting in higher-quality final output frames. To reduce inference complexity, we propose joint inference of RFS and PFE (JISE), achieved through a single execution of STENet for RFS and PFE. Additionally, to effectively manage RFS, PFE, and JISE under RA configuration, we introduce the Space-Time Enhancement Window (STEW), illustrated in Fig.\ref{Random-Access}. For training, we adopt a joint training strategy for STENet, simultaneously optimizing both the synthesis and enhancement pipelines. This simplifies training and thoroughly considers the performance impact between RFS and PFE. The proposed method is integrated into the VVC reference software VTM-15.0, with experimental results showcasing significant performance enhancements in VVC under RA configuration.

The remainder of the paper is organized as follows: Section II reviews the background and related work. Details of our proposed method are given in Section III. Section IV presents our experiments and analysis. Finally, the paper is concluded in Section V.

\section{Related Work}

\subsection{Neural network-based Video Coding Tools}
NNVC tools shall be utilized within traditional coding schemes or together with traditional coding tools. There have been several techniques thoroughly researched, including in-loop filter\cite{Yue-ICIP2021,Wang-CVPRW2022}, post-processing filter\cite{Zhang-ICME2020,Ma-SP2020,Santamaria-ISM2021,Zhang-IEEEAccess2023}, intra prediction\cite{Heming-TMM2020,Dumas-TIP2021}, inter prediction\cite{Zhao-TCSVT2019,Zhao-ICIP2018,Guo-VCIP2020,Hu-ACM2023,Jia-TCSVT2023} and super resolution\cite{Yue-TCSVT2018,Lin-TCSVT2019}. More research about NNVC tools can be found in \cite{Liu-ACM2021, Ma-TCSVT2020}. Meanwhile, JVET has placed significant emphasis on the exploration and standardization of NNVC tools. JVET has established Ad-Hoc Group 11 (AHG11)\cite{Elena-AD0011}, which intends to study the performance, complexity, and training processes of NNVC tools. Some works by JVET experts have achieved significant improvements\cite{Li-arXiv2023}. In addition to well-designed networks and usage strategies, training material needs to include diverse content covering different formats and texture types. JVET NNVC exploration activities have utilized BVI-DVC\cite{Ma-TMM2022} and TVD\cite{Xu-arXiv2021} as training materials. As for neural network inference, F. Galpin et al.\cite{Franck-W0181} propose the Small Ad-hoc Deep-Learning Library (SADL), a header-only small C++ library for the inference of NNVC tools. The objective of SADL is to offer a straightforward framework that is dependency-free and compliant with the software development policy of JVET. 

\subsection{Neural network-based Post-processing Filter Enhancement}
PFE seeks to reduce artifacts introduced by block partitioning and quantization, including block artifacts, ringing artifacts, and over-smoothing. Compared to in-loop filters, PFE does not require any modification to the codec, making it applicable to any video standard. Recently, neural networks have made significant contributions to image and video restoration\cite{Liang-ICCVW2021,Zamir-CVPR2022,Tassano-CVPR2020}, which has catalyzed the vibrant development of neural network-based PFE. Zhang et al.\cite{Zhang-ICME2020} introduce a neural network-based post-processing network for VVC in RA configuration. To enhance the robustness, they trained five different models for five QP groups to accommodate input frames of varying qualities. In \cite{Ma-SP2020}, the MFRNet, comprising four multi-level feature review residual dense blocks (MFRBs), is proposed for post-processing and in-loop filtering. Their experimental results demonstrated that MFRNet is more advantageous for post-processing and in-loop filtering than other popular network structures. Santamaria et al.\cite{Santamaria-ISM2021} propose a content-adaptive convolution neural network post-processing filter, which can be trained offline on general video sequences and later fine-tuned on the test video sequence. 

However, the majority of PFE methods solely rely on the spatial information of the current frame, disregarding the valuable temporal information from neighboring frames. For instance, while the current frame may contain numerous artifacts on the same object, there might be only a few artifacts present in another frame. In such cases, leveraging the temporal information from another frame can help mitigate the artifacts.

\subsection{Neural network-based Reference Frame Synthesis}
RFS aims to synthesize frames that highly similar to the to-be-coded frame to enhance inter prediction. These synthesized frames are inserted into RPLs and treated as the virtual reference frames. Some RFS methods enhance bi-direction prediction by utilizing Video Frame Interpolation (VFI) networks to generate intermediate virtual reference frames. For instance, Zhao et al.\cite{Zhao-ICIP2018} interpolate reconstructed frames as the prediction frame, requiring the encoder to choose between traditional and emerging results using rate-distortion optimization (RDO) at the coding unit (CU) level. Guo et al.\cite{Guo-VCIP2020} synthesize a new reference frame from two-sided previously reconstructed frames and then append the synthesized frame to the RPL, which improves the compression performance of HM-16.20. Furthermore, some studies have explored using video frame extrapolation networks to enhance uni-direction prediction. Huo et al.\cite{Huo-TCSVT2021} utilize a neural network-based extrapolation network to synthesize reference frames, supporting LDP and LDB configurations. Their approach aligns reference frames through block-based motion estimation and compensation, followed by extrapolation from the aligned frames. Jia et al.\cite{Jia-TCSVT2023} present a deep reference frame generation method in which interpolation and extrapolation networks are utilized to enhance bi-directional prediction and uni-directional prediction, respectively. 

Although RFS methods have been developed for many years, there is still room for improvement. Many existing methods simply utilize interpolation or extrapolation networks to directly synthesize reference frames. These approaches often overlook the distortions and artifacts present in the input reconstructed frames, which can negatively impact the quality of the synthesized frames. Furthermore, numerous RFS methods fail to consider the interactions with other NNVC tools. When other NNVC tools are active, the performance of the RFS tool may be significantly hindered\cite{Bao-DRF-AF0208}.

\section{Proposed Method}
\begin{figure*}
\centerline{\includegraphics[width=18cm]{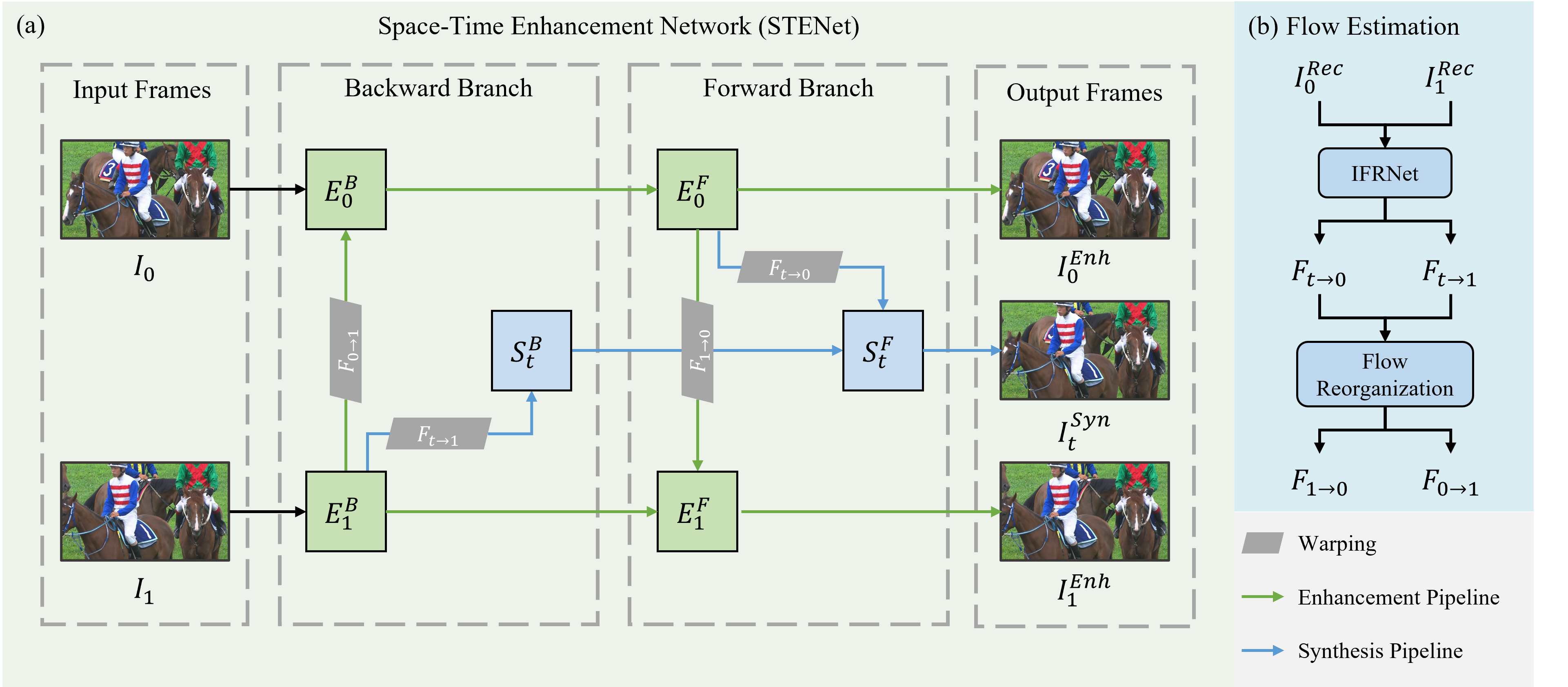}}
\caption{The network architecture of the Space-Time Enhancement Network (STENet). STENet takes two compressed frames with artifacts and distortions, fully leveraging the space-time information to generate an intermediate synthesized frame and two enhanced frames with reduced artifacts. STENet comprises two pipelines: the synthesis pipeline (in blue) for RFS and the enhancement pipeline (in green) for PFE.}
\label{Network Architecture}
\end{figure*}

In this section, we begin by outlining the motivation behind our proposed method. Next, we introduce STENet, as depicted in Fig.\ref{Network Architecture}. Subsequently, we delve into the investigation of RFS, PFE, and JISE within the context of VVC, as illustrated in Fig.\ref{Framework}. Finally, we present the inference details of our proposed method.

\subsection{Motivation}
Despite years of development, there is still room for improvement in NNVC. Initially, most NNVC tools overlook the importance of incorporating space-time information across neighboring frames, eg., PFE methods primarily concentrate on spatial information. Moreover, each NNVC tool is trained independently. Different NNVC tools often require distinct training materials and strategies, leading to substantial human and computational resources. Besides resource consumption, the isolated training of different NNVC tools may result in significant performance overlap. Additionally, there has been limited research exploring the feature reuse among different NNVC tools. By reusing features, different NNVC tools can reduce the complexity of feature extraction and achieve accelerated inference.

To address the aforementioned issues, we propose a joint utilization of RFS and PFE, which contain one joint network, one joint training strategy, and one joint inference. The proposed method realizes the space-time coupled enhancement among neighboring frames. Firstly, we introduce the STENet, which serves the dual purpose of RFS and PFE. The STENet contains a synthesis pipeline that generates a virtual reference frame for RFS, and an enhancement pipeline that improves the quality of two input frames for PFE. Secondly, we adopt a joint training strategy for STENet, with both the synthesis and enhancement pipelines optimized simultaneously. This approach simplifies the training process and demonstrates superior performance compared to the isolated training of two pipelines. Lastly, we propose a joint inference mechanism, JISE, which enables simultaneous inference of RFS and PFE. Unlike independent inference, the JISE significantly reduces the overall inference time by inferring RFS and PFE concurrently, thereby improving efficiency.

\subsection{Space-Time Enhancement Network} \label{STENet paragraph}
The STENet receives two input frames $I_0$ and $I_1$, and produces two enhanced frames $I^{Enh}_0$ and $I^{Enh}_1$ with suppressed artifacts, along with an intermediate synthesized frame $I^{Syn}_t$. This process can be described as:
\begin{equation} I^{Enh}_0, I^{Syn}_t, I^{Enh}_1 = {STENet}(I_{0},I_{1}), \end{equation}
where $STENet$ represents the STENet. As depicted in Fig.\ref{Network Architecture}, the STENet employs a bidirectional recurrent architecture to realize global information propagation. In this bidirectional recurrent setup, both frames and features are propagated in two opposite directions. This allows the input frames to gather information from any other frames, and the intermediate synthesized frame to leverage information from the two input frames. The two recurrent processes are divided into the forward and backward branches depending on the direction of information propagation. 

\subsubsection{Optical Flow Estimation Module}\label{Optical Flow Estimation Module}
The optical flow estimation module is illustrated in Fig.\ref{Network Architecture}(b). To achieve feature alignment between two input frames, it's essential to estimate the optical flows $F_{0 \to 1}$ and $F_{1 \to 0}$ between them. Additionally, for interpolating an intermediate frame from two input frames, estimation of the optical flows $F_{t \to 0}$ and $F_{t \to 1}$ from the intermediate moment $t$ is required. Inspired by \cite{Zhang-TCSVT2023}, which proposes an effective and precise optical flow reusing strategy, we adopt the same strategy to reduce the complexity involved in optical flow estimation.

To begin with, we utilize the IFRNet\cite{Kong-CVPR2022} as the optical flow estimator due to its fast inference speed. Since IFRNet requires inputs in RGB format, we first upsample the U and V channels of each input frame in YUV420 format and then perform a color space transformation to convert them into RGB format. The reformatted input frames are then fed into IFRNet, which estimates the optical flows $F_{t \to 0}$ and $F_{t \to 1}$. This process can be summarized as follows:
\begin{equation}F_{t \to 1}, F_{t \to 0} = \Theta\{I_0, I_1\},\end{equation}
where $\Theta$ represents the intermediate optical flow estimator. Next, following the approach described in \cite{Zhang-TCSVT2023}, we adopt a similar strategy to generate the optical flows $F_{0 \to 1}$ and $F_{1 \to 0}$ based on the previously estimated optical flows $F_{t \to 0}$ and $F_{t \to 1}$. This can be expressed as:
\begin{equation}F_{0 \to 1}, F_{1 \to 0} = \Phi(F_{t \to 0}, F_{t \to 1}),\end{equation}
where $\Phi$ represents the optical flow reusing strategy from \cite{Zhang-TCSVT2023}. At this stage, all the optical flows have been estimated. These resulting optical flows will be utilized for alignment in the bidirectional recurrence, which will be explained in detail in the subsequent section.

\subsubsection{Network Architecture}
Inspired by \cite{Chan-arXiv2022}, we downsample the input frames and the optical flows to reduce the computational cost. For each input frame in YUV420 format, the Y component undergoes a PixelUnshuffle operation\cite{Shi-CVPR2016} with a scale factor of 2, followed by a concatenation operation with the U and V components to generate a six-channel tensor. The optical flows, detailed in Section \ref{Optical Flow Estimation Module}, are downsampled accordingly. These designs enable most of the computations to be performed in the low-resolution feature space, thereby greatly enhancing efficiency. It's worth noting that, instead of downsampling the input frames to estimate optical flows as mentioned in \cite{Chan-arXiv2022},  performing high-resolution optical flow estimation and then downsampling optical flows will yield more accurate low-resolution optical flows.

In the backward branch, information from the succeeding state is transmitted to both the preceding and intermediate states. A weight-shared residual refinement module $\mathcal{R}^B$ is utilized to refine features at each state. The feature extraction process of the succeeding state is described as:
\begin{equation} E_1^B = \mathcal{R}^B ( \mathcal{E}^B ( \mathcal{C}(I_1, Zero) ) ), \end{equation}
where $\mathcal{E}^B$ represents the weight-shared feature extraction module and $\mathcal{C}$ denotes the channel concatenation operation. To align the input channel of $\mathcal{E}^B$ in each state, the tensor $Zero$ filled with zeros is treated as a placeholder. Then, the preceding state receives knowledge from the succeeding feature:
\begin{equation} E_0^B = \mathcal{R}^B(\mathcal{E}^B ( \mathcal{C}(I_0, \mathcal{W}(E_1^B, F_{0 \to 1}) ) ) ), \end{equation}
where $\mathcal{W}$ denotes the flow warping operation. Next, the knowledge from the succeeding state is propagated to the intermediate state as prior knowledge for the frame synthesis, represented as:
\begin{equation} S_t^B = \mathcal{R}^B(\mathcal{W}(E_1^B, F_{t \to 1}) ). \end{equation}
Up to now, via backward propagation, we acquire three features in distinct states, which are intended for further refinement within the forward branch.

The forward branch shares the same idea with the backward branch. Both the succeeding and intermediate states will receive knowledge from the preceding state. The specific processes at three different states are as follows:
\begin{equation} E_0^F = \mathcal{R}^F (\mathcal{C}(E_0^B, I_0^{Rec}, Zero)), \end{equation}
\begin{equation} E_1^F = \mathcal{R}^F (\mathcal{C}(E_1^B, \mathcal{W}(E_0^F, F_{1 \to 0}))), \end{equation}
\begin{equation} S_t^F = \mathcal{R}^F (\mathcal{C}(S_t^B, \mathcal{W}(E_0^F, F_{t \to 0}))), \end{equation}
where \(\mathcal{R}^F\) represents the weight-shared residual refinement module in the forward branch. So far, we have extracted features in three states. A weight-shared convolution layer was employed for each state for reconstruction, yielding a six-channel tensor. To get the output frame in YUV420 format, the first four channels of the reconstructed feature undergo a PixelShuffle operation\cite{Shi-CVPR2016} with a scale factor of $2$ to obtain the Y component, and the last two channels are U and V components, respectively.

As depicted in Fig.\ref{Network Architecture}(a), STENet comprises two pipelines: the enhancement pipeline and the synthesis pipeline. These pipelines can operate jointly or independently based on routing selection. The enhancement pipeline takes two input frames $I_0$ and $I_1$ as input to enhance each other, resulting in two enhanced frames $I^{Enh}_0$ and $I^{Enh}_1$:
\begin{equation} I^{Enh}_0, I^{Enh}_1 = {Enh}(I_{0},I_{1}) , \end{equation}
where the $Enh$ represents the enhancement pipeline. Meanwhile, the synthesis pipeline takes two input frames $I_0$ and $I_1$ for frame synthesis, producing an intermediate synthesized frame $I^{Syn}_t$:
\begin{equation} I^{Syn}_t = {Syn}(I_{0},I_{1}) , \end{equation}
where the $Syn$ denotes the synthesis pipeline. Additionally, the enhancement pipeline can be utilized for PFE, the synthesis pipeline for RFS, and the entire STENet for JISE. Further details are provided in the subsequent section.

\subsection{Integration into VVC}

\begin{figure*}
\centerline{\includegraphics[width=18cm]{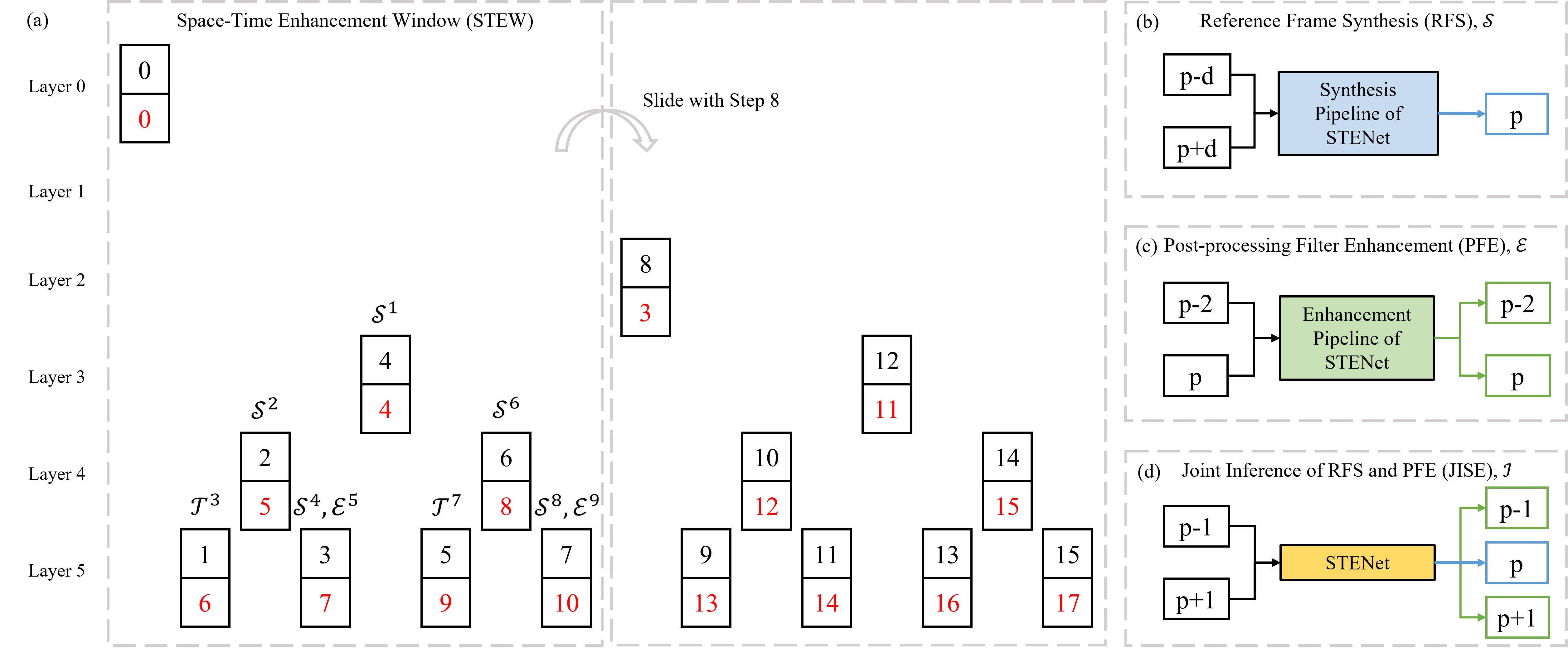}}
\caption{The Space-Time Enhancement Window (STEW) is proposed to manage RFS, PFE, and JISE effectively under RA configuration. Each STEW contains 8 consecutive frames. Within the current STEW, we determine whether to execute RFS, PFE, or JISE based on the POC $p$ of the current frame. Once the last frame is decoded and enhanced, the STEW will slide forward by 8 POC distances.}
\label{Random-Access}
\end{figure*}

The framework of our proposed method, as illustrated in Fig.\ref{Framework}, integrates both RFS and PFE into the VVC reference software VTM-15.0. RFS aims to synthesize frames that highly similar to the to-be-coded frame to enhance inter prediction. These synthesized frames are inserted into RPLs and treated as the virtual reference frames. On the other hand, PFE operates outside the encoding and decoding loops to enhance the reconstructed frames. Meanwhile, to reduce computation costs, we propose JISE, in which the RFS and PFE are executed jointly.

\subsubsection{Space-Time Enhancement Window}
To effectively coordinate RFS, PFE, and JISE under RA configuration, we introduce the STEW. As depicted in Fig.\ref{Random-Access}(a), each STEW contains 8 consecutive frames. Within the current STEW, we determine whether to execute RFS, PFE, or JISE based on the POC $p$ of the current to-be-coded frame. The specific mode decision process is as follows:
\begin{equation}
mode =
\begin{cases}
\mathcal{S} \quad (p \bmod 8) \in \{2,3,4,6,7\} \\
\mathcal{E} \quad (p \bmod 8) \in \{3,7\} \\
\mathcal{J} \quad (p \bmod 8) \in \{1,5\} \\
\end{cases}
,
\end{equation}
where $\mathcal{S}$ denotes RFS, $\mathcal{E}$ represents PFE, and $\mathcal{J}$ indicates JISE. 

In each STEW, the processes of RFS, PFE, and JISE are executed alternately, their execution sequence is constrained by the encoding order of the RA configuration. As illustrated in Fig.\ref{Random-Access}(a), we annotate their execution sequence using symbols. For a specific symbol, the superscript denotes its order, while the subscript indicates whether it is RFS, PFE, or JISE. $\mathcal{S}^1$ synthesizes an additional reference frame for the $4^{th}$ frame, followed by $\mathcal{S}^2$, which synthesizes an additional reference frame for the $2^{nd}$ frame. Subsequently, $\mathcal{J}^3$ synthesizes an additional reference frame for the $1^{st}$ frame and enhances the $0^{th}$ and $2^{nd}$ frames. Finally, $\mathcal{S}^4$ synthesizes an additional reference frame for the $3^{rd}$ frame, and $\mathcal{E}^5$ enhances the $1^{st}$ and $3^{rd}$ frames. Further operations are not described in detail. In each STEW, the latter 7 frames will reference the synthesized reference frames generated by RFS or JISE, and all frames will be enhanced by PFE or JISE. Once the last frame is decoded and enhanced, the STEW will slide forward by 8 POC distances.

\subsubsection{Reference Frame Synthesis}
DPB stores the reconstructed frames during encoding. Specific reconstructed frames are selected from DPB to be included in the RPL as reference frames. For bi-directional inter prediction, two separate RPLs are established in different directions for the to-be-coded frame. The encoder searches the best-matching blocks in reference frames as predictions of the current to-be-coded CU. Reference frames with more similar content to the to-be-coded frame can lead to smaller residual coding consumption in the bitstream. To enhance the bi-directional inter prediction in VVC, the JVET CTC \cite{Bossen-CTC-Y2010} specifies RA configuration, which employs a hierarchical coding structure with 32 frames in a group of pictures (GOP).

Within each STEW, RFS synthesizes a virtual reference frame to enhance bi-directional inter prediction under RA configuration. As illustrated in Fig.\ref{Random-Access}(a), the frames within a GOP are divided into 6 temporal layers in RA configuration, allowing the utilization of previously encoded frames in lower layers as references for the to-be-coded frames in higher layers. As shown in Fig.\ref{Random-Access}(b), RFS selects two-sided reconstructed frames $I_{p-d}^{Rec}$ and $I_{p+d}^{Rec}$ as inputs to synthesize a virtual reference frame $I_{p}^{Syn}$ through the STENet's synthesis pipeline, represented as:
\begin{equation}
I^{Syn}_p = {Syn}(I_{p-d}^{Rec},I_{p+d}^{Rec}) ,
\end{equation}
where $Syn$ represents STENet's synthesis pipeline. Considering the degradation of frame synthesis performance over long temporal distances, the POC distance $d$ should be less than 4. In current STEW, the relationship between the POC distance $d$ and the to-be-coded frame's POC $p$ is arranged as:
\begin{equation}
d =
\begin{cases}
    4 \quad (p \bmod 8) = 4 \\
    2 \quad (p \bmod 8) \in \{2,6\} \\
    1 \quad (p \bmod 8) \in \{3,7\} \\
\end{cases}
.
\end{equation}
As for the other frames, in which $(p \bmod 8) \in \{1,5\}$, we will provide detailed explanations in Section \ref{Joint-RFS-and-PFE}.

The synthesized frame is inserted into two RPLs (RPL0 and RPL1) of the current to-be-coded frame and treated as additional reference frames in two RPLs. The indexes of the synthesized frame in RPL0 and RPL1 are set to 1. Notably, the synthesized reference frame is utilized like other reference frames, thus ensuring that the encoder upholds consistent ME and MC procedures during inter prediction. RFS operates in both the encoding and decoding processes, eliminating the necessity to convey additional information into the bitstream.

\subsubsection{Post-processing Filter Enhancement} \label{Post-processing Filter Enhancement}
In block-based hybrid coding schemes, reconstructed frames often suffer from block artifacts, ringing artifacts, and over-smoothing. Those artifacts will degrade the perceptual quality of the reconstructed video. In most block-based hybrid coding schemes, several in-loop filters operate inside the decoder to alleviate artifacts in filtered frames. On the other hand, post-processing filters operate out of the encoding and decoding loop. This means that post-processing filters can be integrated into any video standard without any codec modifications. However, most post-processing filters only use information from the current to-be-coded frame, ignoring potentially useful information from neighboring frames.

Within each STEW, PFE filters the reconstructed frames to mitigate their artifacts and distortions. PFE simultaneously improves two input frames, enabling each frame to leverage its own spatial information and the temporal information from another frame to improve its quality. As shown in Figure \ref{Random-Access}(c), PFE takes two reconstructed frames $I^{Rec}_{p-d}$ and $I^{Rec}_{p}$ as inputs and uses the STENet's enhancement pipeline to improve them. If the POC distance $d$ between the two reconstructed frames is either too short or too long, the reconstructed frame will not be able to effectively get complementary information from other frames. Therefore, we set the POC distance $d$ between the two reconstructed frames to be 2. The process can be described as follows:
\begin{equation}
I^{Enh}_{p-2}, I^{Enh}_{p}= {Enh}(I_{p-d}^{Rec},I_{p}^{Rec}) \quad (p \bmod 8) \in \{3,7\},
\end{equation}
where $Enh$ represents STENet's enhancement pipeline, $I^{Enh}_{p-d}$ and $I^{Enh}_{p}$ are two enhanced frames, and $p$ is the POC of the current to-be-coded frame. In other words, the $1^{st}$ and $3^{rd}$ reconstructed frames are filtered simultaneously, while the $5^{th}$ and $7^{th}$ reconstructed frames are filtered simultaneously. As for the other frames in the current STEW, we will provide detailed explanations in Section \ref{Joint-RFS-and-PFE}. The enhanced frames are considered as the final output frames and are not used as reference frames.

If the reconstructed frames exhibit minimal artifacts, PFE may degrade the quality of the enhanced frames. Hence, we offer the flexibility to enable or disable PFE at the block level. We utilize PSNR metrics to assess the similarity between the original, filtered, and reconstructed blocks. If the original block closely resembles the filtered block more than the reconstructed block, we activate PFE for that block. In contrast, we will turn off the PFE in this block. Switch flags indicating the PFE status are subsequently embedded into the bitstream and transmitted to the decoder.

\subsubsection{Joint Inference of RFS and PFE}\label{Joint-RFS-and-PFE}
In our design, the PFE and RFS are continuous in some positions. For instance, once the $0^{th}$ and $2^{nd}$ frames in the current STEW are decoded, PFE can utilize STENet's enhancement pipeline to filter them. After PFE, RFS can utilize STENet's synthesis pipeline to synthesize a virtual reference frame for the $1^{st}$ frame. This inspires us to introduce the JISE, in which RFS and PFE are executed jointly. Instead of utilizing RFS and PFE independently, JISE can reduce network inference complexity by reusing features between two pipelines of STENet.

For the first three frames in the current STEW, we can input the  $0^{th}$ and $2^{nd}$  reconstructed frames into STENet to filter them and synthesize an intermediate frame. The synthesized frame is treated as the virtual reference frame for the $1^{st}$ frame. As for the $4^{th}$, $5^{th}$, and $6^{th}$ frames in the current STEW, the process is the same. The process can be described as follows:
\begin{equation}
\begin{split}
& I^{Enh}_{p-1}, I^{Syn}_p , I^{Enh}_{p+1} = {STENet}(I_{p-1}^{Rec},I_{p+1}^{Rec}) \\
& \quad (p \bmod 8) \in \{1,5\}, 
\end{split}
\end{equation}
where $STENet$ represents the STENet. The enhanced frames $I^{Enh}_{p-1}$ and $I^{Enh}_{p+1}$ are treated as final output frames after determining whether to keep enhanced blocks or not at the block level. The synthesized frame $I^{Syn}_p$ is treated as a virtual reference frame, which will be inserted into two RPLs of the current to-be-coded frame $I_p$.

The synthesis and enhancement pipelines reuse lots of modules. When RFS and PFE are executed independently, those reused modules will be inferred twice. In contrast, those reused modules will infer only once within JISE. That is why the proposed JISE can reduce network inference complexity.

\subsection{Inference Details}

\subsubsection{Block-based Network Inference}
During network inference, large input sizes can lead to excessive memory consumption, especially for $3840 \times 2160$ (4K) resolution inputs. To solve this problem, we divide the input frames into blocks and work on them separately. However, using small blocks can limit the amount of spatial and temporal information captured, leading to a notable drop in STENet performance. Therefore, we use different block sizes for different resolutions. For JVET NNVC common test condition (CTC) class D\cite{Elena-CTC-AE2016} input frames, we use $240 \times 240$ blocks with an extra padding of $8$. For input frames from other CTC classes, we use $480 \times 480$ blocks with an extra padding of $16$. It is worth mentioning that the activation or deactivation decisions for the PFE are also made at the block level as previously described.

\subsubsection{Temporary Buffer for Enhanced Frames}
After PFE, the enhanced frames are temporarily stored in a buffer and released as output frames once all frames within the current STEW are enhanced. The temporary buffer ensures that the inputs of RFS must be the reconstructed frames without being enhanced by PFE.

Without this design, frames enhanced by PFE will be utilized for subsequent RFS. For example, as depicted in Fig.\ref{Random-Access}(a), following the $\mathcal{J}^3$ process, the $2^{nd}$ frame is enhanced. This enhanced $2^{nd}$ frame is subsequently utilized as input during the $\mathcal{S}^4$ process, potentially degrading RFS performance. Specifically, during training, STENet takes reconstructed frames with compression artifacts as inputs. However, during inference, if the inputs of STENet are enhanced frames with suppressed artifacts, the quality of the synthesized frame may decline. Moreover, the synthesized frame may not serve as an effective reference for the to-be-coded frame.

\section{Experiments}
This section provides a detailed elucidation of the experimental setup and results. It includes details about the training data generation and the joint training strategy. Furthermore, we present both subjective and objective analyses of the proposed method, along with ablation experiments. Additionally, we conduct comparisons between the proposed method and existing methodologies.

\begin{table*}
\caption{Performance of the Proposed Method Compared with VTM-15.0 under RA configuration}
\label{VTM-BDrate}
\setlength{\tabcolsep}{10pt}
\renewcommand\arraystretch{1.2}
\begin{center}
\begin{tabular}{cc|cccccc}
\hline
\multicolumn{1}{c|}{\multirow{2}{*}{\textbf{Class}}} & \multirow{2}{*}{\textbf{Sequence}} & \multicolumn{6}{c}{\textbf{BD-rate}}                                                                                                                      \\ \cline{3-8} 
\multicolumn{1}{c|}{}                                &                                    & \textbf{Y-PSNR}       & \textbf{U-PSNR}        & \multicolumn{1}{c|}{\textbf{V-PSNR}}        & \textbf{Y-MSIM}       & \textbf{U-MSIM}        & \multicolumn{1}{c}{\textbf{V-MSIM}}              \\ \hline
\multicolumn{1}{c|}{\multirow{4}{*}{Class A1}}       & Tango2                             & -10.80\%          & -25.75\%          & \multicolumn{1}{c|}{-24.47\%}          & -11.91\%                & -29.64\%                 & \multicolumn{1}{c}{-27.15\%}                \\
\multicolumn{1}{c|}{}                                & FoodMarket4                        & -4.88\%          & -10.89\%           & \multicolumn{1}{c|}{-11.76\%}           & -5.43\%                & -12.86\%                 & \multicolumn{1}{c}{-13.42\%}                \\
\multicolumn{1}{c|}{}                                & Campfire                           & -2.17\%          & -0.02\%           & \multicolumn{1}{c|}{-4.72\%}           & -3.73\%                & -1.06\%                 & \multicolumn{1}{c}{-8.47\%}                \\ \cline{2-8} 
\multicolumn{1}{c|}{}                                & Average                            & -5.95\%          & -12.22\%           & \multicolumn{1}{c|}{-13.65\%}           & -7.02\%                & -14.52\%                 & \multicolumn{1}{c}{-16.35\%}                \\ \hline
\multicolumn{1}{c|}{\multirow{4}{*}{Class A2}}       & CatRobot                           & -9.17\%          & -23.55\%          & \multicolumn{1}{c|}{-19.90\%}          & -10.04\%                & -20.54\%                 & \multicolumn{1}{c}{-16.87\%}                \\
\multicolumn{1}{c|}{}                                & DaylightRoad2                      & -10.42\%          & -19.19\%          & \multicolumn{1}{c|}{-13.69\%}          & -10.30\%                & -20.59\%                 & \multicolumn{1}{c}{-14.43\%}                \\
\multicolumn{1}{c|}{}                                & ParkRunning3                       & -2.83\%          & -4.49\%           & \multicolumn{1}{c|}{-3.19\%}           & -4.65\%                & -6.74\%                 & \multicolumn{1}{c}{-5.36\%}                \\ \cline{2-8} 
\multicolumn{1}{c|}{}                                & Average                            & -7.48\%          & -15.74\%          & \multicolumn{1}{c|}{-12.26\%}           & -8.33\%                & -15.96\%                 & \multicolumn{1}{c}{-12.22\%}                \\ \hline
\multicolumn{1}{c|}{\multirow{6}{*}{Class B}}        & MarketPlace                        & -5.75\%          & -19.58\%          & \multicolumn{1}{c|}{-16.86\%}          & -6.33\%          & -20.71\%          & \multicolumn{1}{c}{-15.65\%}          \\
\multicolumn{1}{c|}{}                                & RitualDance                        & -6.60\%          & -13.23\%           & \multicolumn{1}{c|}{-17.44\%}           & -6.73\%          & -14.22\%           & \multicolumn{1}{c}{-17.97\%}          \\
\multicolumn{1}{c|}{}                                & Cactus                             & -6.25\%          & -15.28\%          & \multicolumn{1}{c|}{-8.08\%}           & -7.15\%          & -18.09\%          & \multicolumn{1}{c}{-12.81\%}         \\
\multicolumn{1}{c|}{}                                & BasketballDrive                    & -7.67\%          & -17.90\%          & \multicolumn{1}{c|}{-15.42\%}          & -8.73\%          & -19.65\%           & \multicolumn{1}{c}{-21.36\%}         \\
\multicolumn{1}{c|}{}                                & BQTerrace                          & -5.37\%          & -15.87\%          & \multicolumn{1}{c|}{-11.91\%}           & -5.47\%          & -16.54\%           & \multicolumn{1}{c}{-14.46\%}           \\ \cline{2-8} 
\multicolumn{1}{c|}{}                                & Average                            & -6.33\%          & -16.37\%          & \multicolumn{1}{c|}{-13.94\%}           & -6.88\%          & -17.84\%          & \multicolumn{1}{c}{-16.63\%}          \\ \hline
\multicolumn{1}{c|}{\multirow{5}{*}{Class C}}        & BasketballDrill                    & -7.92\%          & -16.88\%          & \multicolumn{1}{c|}{-17.18\%}           & -9.97\%          & -18.48\%          & \multicolumn{1}{c}{-21.31\%}         \\
\multicolumn{1}{c|}{}                                & BQMall                             & -10.61\%          & -26.10\%          & \multicolumn{1}{c|}{-25.79\%}          & -12.62\%          & -25.81\%          & \multicolumn{1}{c}{-24.69\%}         \\
\multicolumn{1}{c|}{}                                & PartyScene                         & -6.39\%          & -14.89\%          & \multicolumn{1}{c|}{-13.43\%}          & -6.25\%          & -12.84\%          & \multicolumn{1}{c}{-12.16\%}         \\
\multicolumn{1}{c|}{}                                & RaceHorses                         & -7.59\%          & -23.38\%          & \multicolumn{1}{c|}{-26.08\%}          & -9.54\%          & -21.43\%          & \multicolumn{1}{c}{-23.31\%}         \\ \cline{2-8} 
\multicolumn{1}{c|}{}                                & Average                            & -8.13\%          & -20.31\%          & \multicolumn{1}{c|}{-20.62\%}          & -9.59\%          & -19.64\%          & \multicolumn{1}{c}{-20.37\%}         \\ \hline
\multicolumn{1}{c|}{\multirow{5}{*}{Class D}}        & BasketballPass                     & -10.00\%          & -26.44\%          & \multicolumn{1}{c|}{-23.94\%}          & -10.53\%          & -28.54\%          & \multicolumn{1}{c}{-24.76\%}         \\
\multicolumn{1}{c|}{}                                & BQSquare                           & -10.54\%          & -13.39\%          & \multicolumn{1}{c|}{-21.89\%}          & -7.35\%          & -15.06\%           & \multicolumn{1}{c}{-19.11\%}          \\
\multicolumn{1}{c|}{}                                & BlowingBubbles                     & -5.33\%          & -14.40\%          & \multicolumn{1}{c|}{-13.41\%}          & -5.00\%          & -14.02\%          & \multicolumn{1}{c}{-12.75\%}         \\
\multicolumn{1}{c|}{}                                & RaceHorses                         & -9.20\%          & -25.79\%          & \multicolumn{1}{c|}{-27.27\%}          & -9.24\%          & -25.53\%          & \multicolumn{1}{c}{-24.35\%}         \\ \cline{2-8} 
\multicolumn{1}{c|}{}                                & Average                            & -8.77\%          & -20.00\%          & \multicolumn{1}{c|}{-21.63\%}          & -8.03\%          & -20.29\%          & \multicolumn{1}{c}{-20.24\%}         \\ \hline
\multicolumn{2}{c|}{\textbf{Average}}                                                     & \textbf{-7.34\%} & \textbf{-17.21\%} & \multicolumn{1}{c|}{\textbf{-16.65\%}} & \textbf{-7.95\%} & \textbf{-17.91\%} & \multicolumn{1}{c}{\textbf{-17.44\%}} \\ \hline
\end{tabular}
\end{center}
\end{table*}

\subsection{Training Details}
\subsubsection{Training Data}
We adopt the Vimeo-90k triplet dataset\cite{Xue-IJCV2019} for training. The Vimeo-90k dataset consists of 73,171 3-frame sequences with a fixed resolution of $448 \times 256$. During inference, the inputs for STENet consist of two reconstructed frames containing compression artifacts, while the outputs consist of an intermediate synthesized frame and two enhanced frames with suppressed artifacts. Compression artifacts are introduced into the training data to simulate real-world inference scenarios. Specifically, we compress the first and third frames in each triplet using the all intra (AI) configuration, while employing three consecutive original frames as ground-truth frames. The quantization parameter (QP) for each triplet is randomly selected from 22, 27, 32, 37, and 42.

\subsubsection{Training Strategy}\label{Training-Strategy}
The STENet is trained using PyTorch on four Nvidia Geforce RTX 3060 GPUs. We use a Charbonnier penalty function\cite{Lai-CVPR2017} as the loss function and supervise all three output frames of STENet simultaneously:
\begin{equation} 
L = \sqrt{ {\left \| {I^{Ste} - I^{GT}} \right \|}^2 + {\epsilon}^2 } 
\end{equation}
where $I^{Ste}$ is the output frames of STENet and $I^{GT}$ denotes ground-truth frames. $\epsilon$ is set to $10^{-3}$ as a constant. The loss function is optimized through the Adam optimizer with ${\beta}_{1} = 0.9$ and ${\beta}_{2} = 0.999$. Before training, the STENet loads the pre-trained weights of IFRNet. The initial learning rates of IFRNet and other parts are set to $2.5 \times 10^{-5}$ and $1 \times 10^{-4}$ respectively and decay to $10^{-7}$ with one cosine annealing\cite{Loshchilov-ICLR2017}. For data augmentation, two input frames are randomly cropped into $128 \times 128$ patches, and their order is randomly swapped. The batch size is 48, and the training lasts 300 epochs.

\subsection{Experimental Results and Analysis}

\begin{table}
\caption{Bitrate Reduction and Y-PSNR Improvement Compared with VTM-15.0 under RA Configuration}
\label{VTM-Bitrate-PSNR}
\setlength{\tabcolsep}{7pt}
\renewcommand\arraystretch{1.2}
\begin{center}
\begin{tabular}{c|cc}
\hline
\multirow{1}{*}{\textbf{Class}}   & \textbf{Bitrate Reduction}      & \textbf{Y-PSNR Improvement}  \\ \hline
Class A1                          & -2.77\%                         & 0.19\%          \\
Class A2                          & -4.04\%                         & 0.18\%          \\
Class B                           & -3.28\%                         & 0.24\%          \\
Class C                           & -5.29\%                         & 0.31\%          \\
Class D                           & -4.90\%                         & 0.48\%          \\ \hline
\textbf{Average}                  & \textbf{-4.08\%}                & \textbf{0.29\%} \\ \hline
\end{tabular}
\end{center}
\end{table}

\begin{table}
\caption{Computational Time Compared with VTM-15.0 under RA Configuration}
\label{VTM-Time}
\setlength{\tabcolsep}{16pt}
\renewcommand\arraystretch{1.2}
\begin{center}
\begin{tabular}{c|cccccc}
\hline
\multirow{2}{*}{\textbf{Class}} & \multicolumn{2}{c}{\textbf{Computational Time (CPU Only)}} \\ \cline{2-3}
                                & \textbf{Encoding} & \multicolumn{1}{c}{\textbf{Decoding}} \\ \hline
Class A1                        & 274\%             & \multicolumn{1}{c}{270045\%}    \\
Class A2                        & 256\%             & \multicolumn{1}{c}{247660\%}    \\
Class B                         & 302\%             & \multicolumn{1}{c}{335021\%}   \\
Class C                         & 197\%             & \multicolumn{1}{c}{208032\%}    \\
Class D                         & 190\%             & \multicolumn{1}{c}{180358\%}    \\ \hline
\textbf{Average}                & \textbf{272\%}    & \multicolumn{1}{c}{\textbf{261266\%}}  \\ \hline
\end{tabular}
\end{center}
\end{table}

The proposed method is integrated into the VVC reference software VTM-15.0. The performance of our method is evaluated using VTM-15.0 as the benchmark anchor under RA configuration, with the compression performance measured by the Bjøntegaard Delta rate (BD-rate)\cite{Gisle-BDRate}. A negative (positive) BD-rate percentage value indicates an increase (decrease) in compression efficiency performance. Follow the JVET NNVC CTC\cite{Elena-CTC-AE2016}, each sequence is encoded with five QP, including 22, 27, 32, 37, and 42. Table \ref{VTM-BDrate} reports the PSNR-based BD-Rate and the MSIM-based BD-Rate for each sequence. As can be observed, the average PSNR-based BD-Rate and average MSIM-based BD-Rate are -7.34\%/-17.21\%/-16.65\% and -7.95\%/-17.91\%/-17.44\% respectively.

\begin{figure*}[htbp]
\begin{center}
    \begin{minipage}[t]{0.26\textwidth}
        \centering
        \includegraphics[height=0.5\textwidth]{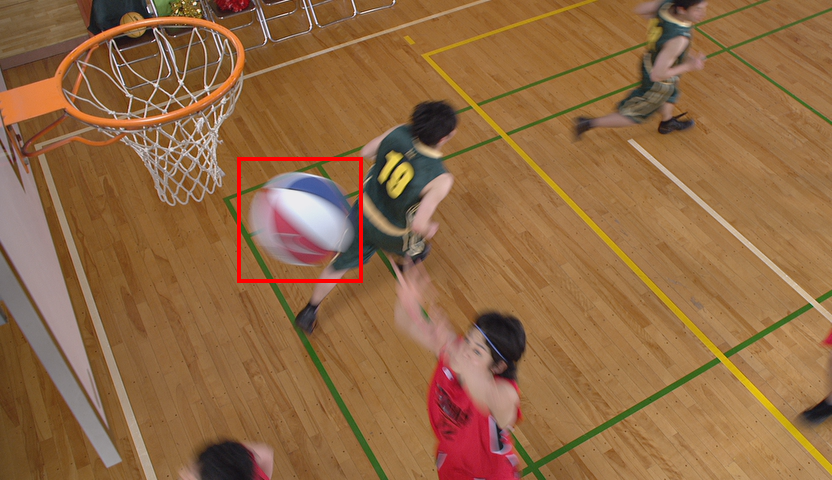} \\
        \vspace{0.02\textwidth}
        \includegraphics[height=0.5\textwidth]{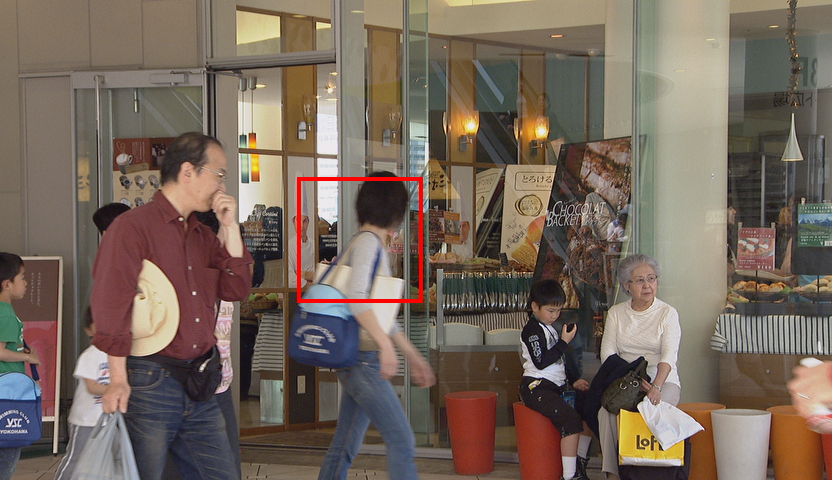} \\
        \vspace{0.02\textwidth}
        \includegraphics[height=0.5\textwidth]{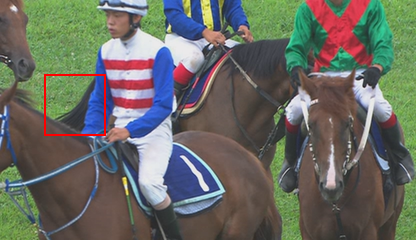} \\
        \vspace{0.02\textwidth}
        \centerline{\footnotesize Original}\medskip
    \end{minipage}
    \hspace{-4mm}
    \begin{minipage}[t]{0.13\textwidth}
        \centering
        \includegraphics[height=\textwidth]{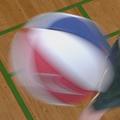} \\
        \vspace{0.04\textwidth}
        \includegraphics[height=\textwidth]{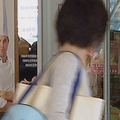} \\
        \vspace{0.04\textwidth}
        \includegraphics[height=\textwidth]{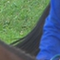} \\
        \vspace{0.04\textwidth}
        \centerline{\footnotesize Original Crop}\medskip
    \end{minipage}
    \begin{minipage}[t]{0.13\textwidth}
        \centering
        \includegraphics[height=\textwidth]{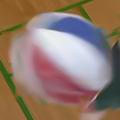} \\
        \vspace{0.04\textwidth}
        \includegraphics[height=\textwidth]{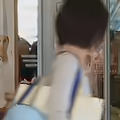} \\
        \vspace{0.04\textwidth}
        \includegraphics[height=\textwidth]{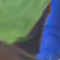} \\
        \vspace{0.04\textwidth}
        \centerline{\footnotesize VTM-15.0 - QP37}\medskip
    \end{minipage}
    \begin{minipage}[t]{0.13\textwidth}
        \centering
        \includegraphics[height=\textwidth]{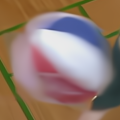} \\
        \vspace{0.04\textwidth}
        \includegraphics[height=\textwidth]{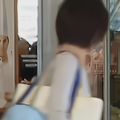} \\
        \vspace{0.04\textwidth}
        \includegraphics[height=\textwidth]{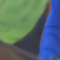} \\
        \vspace{0.04\textwidth}
        \centerline{\footnotesize Ours - QP37}\medskip
    \end{minipage}
    \begin{minipage}[t]{0.13\textwidth}
        \centering
        \includegraphics[height=\textwidth]{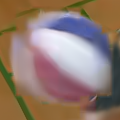} \\
        \vspace{0.04\textwidth}
        \includegraphics[height=\textwidth]{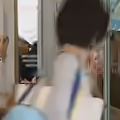} \\
        \vspace{0.04\textwidth}
        \includegraphics[height=\textwidth]{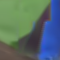} \\
        \vspace{0.04\textwidth}
        \centerline{\footnotesize VTM-15.0 - QP42}\medskip
    \end{minipage}
    \begin{minipage}[t]{0.13\textwidth}
        \centering
        \includegraphics[height=\textwidth]{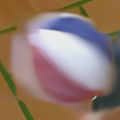} \\
        \vspace{0.04\textwidth}
        \includegraphics[height=\textwidth]{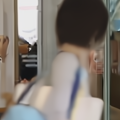} \\
        \vspace{0.04\textwidth}
        \includegraphics[height=\textwidth]{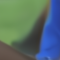} \\
        \vspace{0.04\textwidth}
        \centerline{\footnotesize Ours - QP42}\medskip
    \end{minipage}
    \caption{Visual comparisons between the reconstructed frames generated by the anchor VTM-15.0 and our proposed method. The visual comparisons are conducted under two QP values (37 and 42). The top row showcases the $9^{th}$  frame of \emph{BasketballDrill}, the middle row displays the $9^{th}$ frame of \emph{BQMall}, and the bottom row features the $9^{th}$ frame of \emph{RaceHorses}.}
    \label{Visual-Comparison}
\end{center}
\end{figure*}

The visual comparisons of the reconstructed frames between the anchor VTM-15.0 and our proposed method are depicted in Fig. \ref{Visual-Comparison}. We use two high QP values, 37 and 42, to better observe visual differences. It can be observed that the anchor fails to handle the ME and MC of moving objects, such as the basketball's edge and the woman's back. Additionally, noticeable blocking and ringing artifacts persist in the reconstructed frames of the anchor, for instance, in the horse's tail. In contrast, the reconstructed frames produced by our method exhibit higher quality. For example, the basketball edge appears smoother, the woman's back is distinct, and the horse's tail is more continuous. Our method effectively alleviates the issues present in the anchor. Specifically, RFS generates virtual reference frames to provide more references for the current to-be-coded frame, while PFE filters the reconstructed frames to eliminate artifacts.

\begin{figure*}
\begin{center}
\subfigure[Class A1 - Tango2]{
\includegraphics[width=8.3cm]{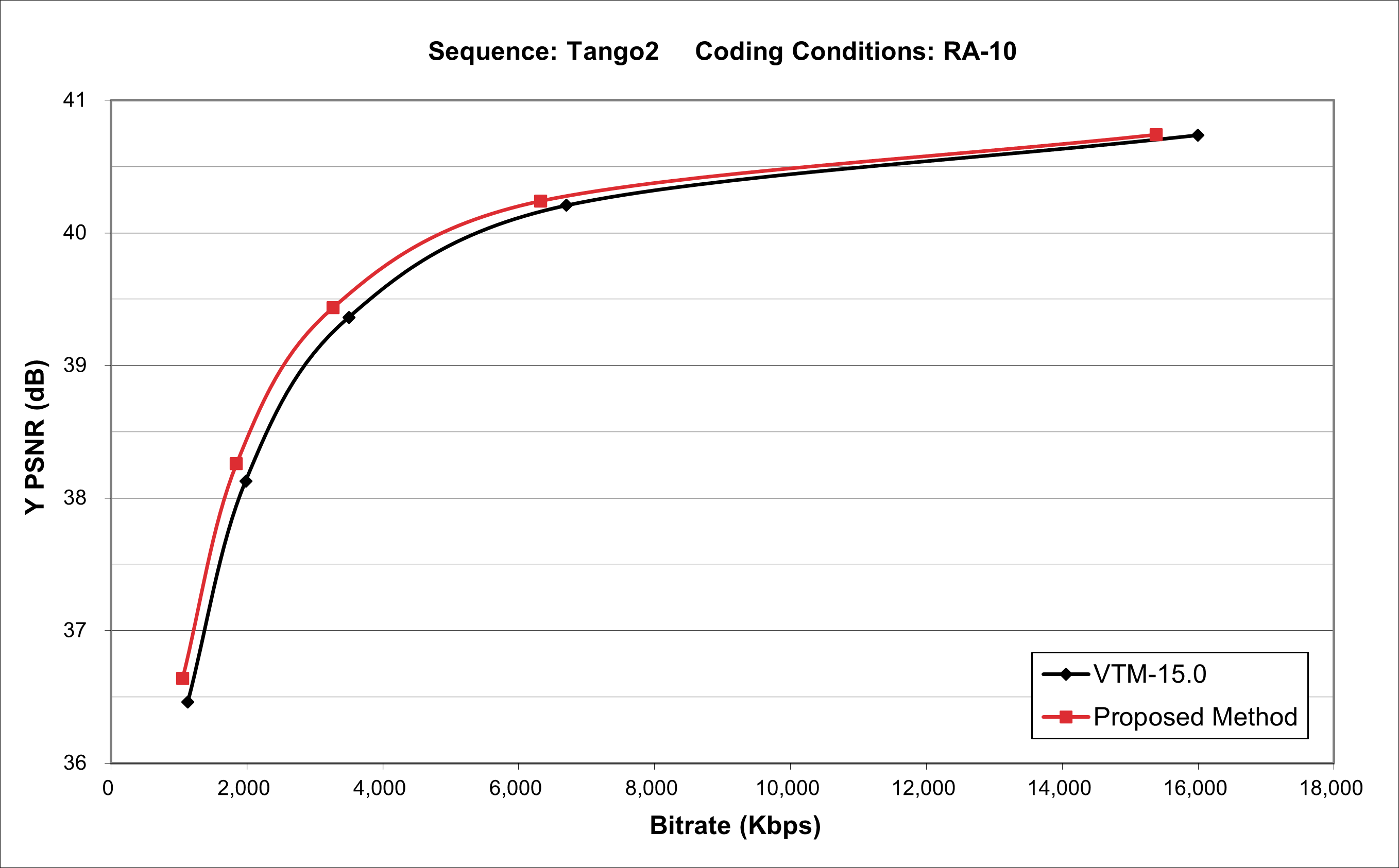}}
\quad
\subfigure[Class B - BasketballDrive]{
\includegraphics[width=8.3cm]{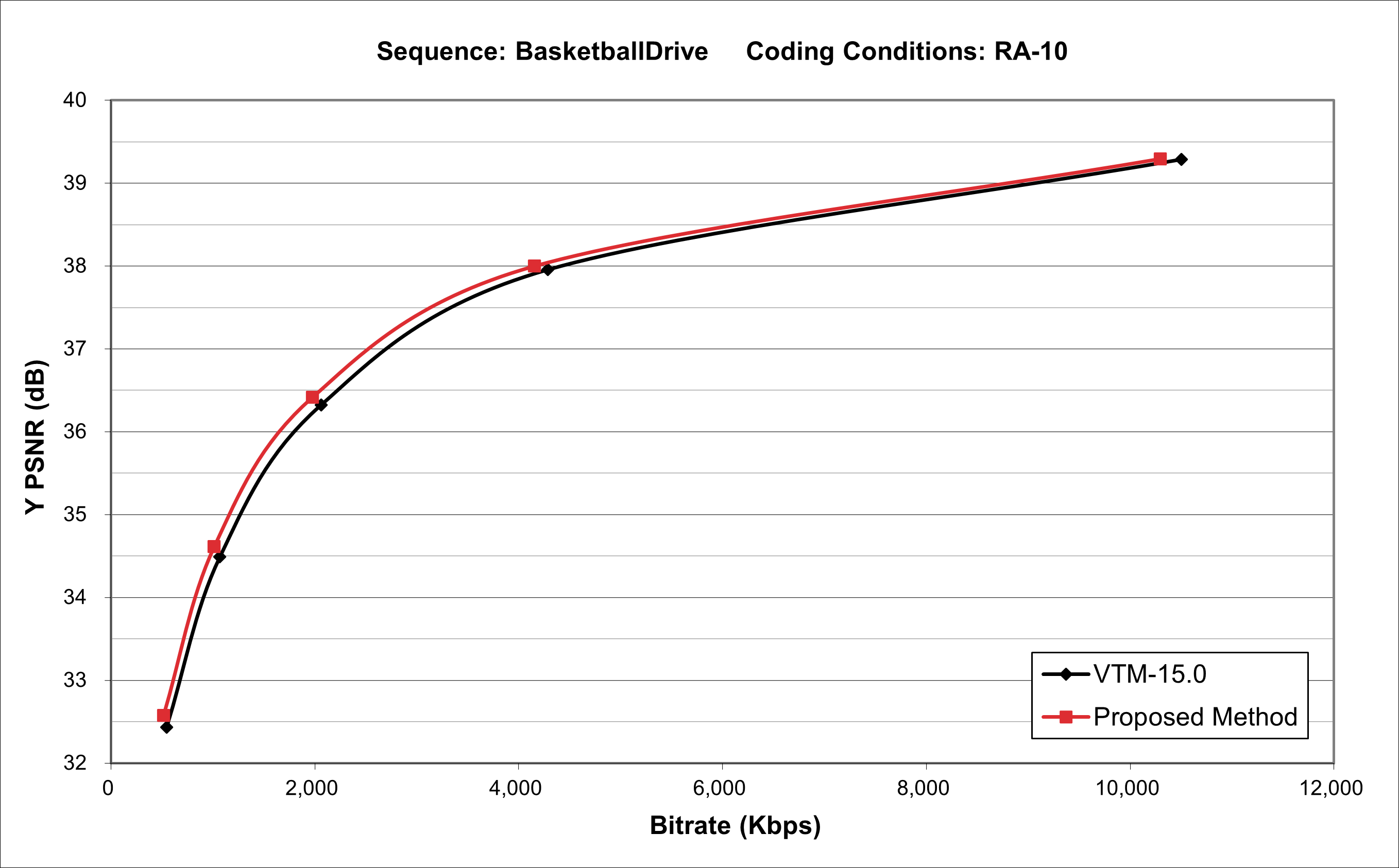}}
\quad
\subfigure[Class C - BQMall]{
\includegraphics[width=8.3cm]{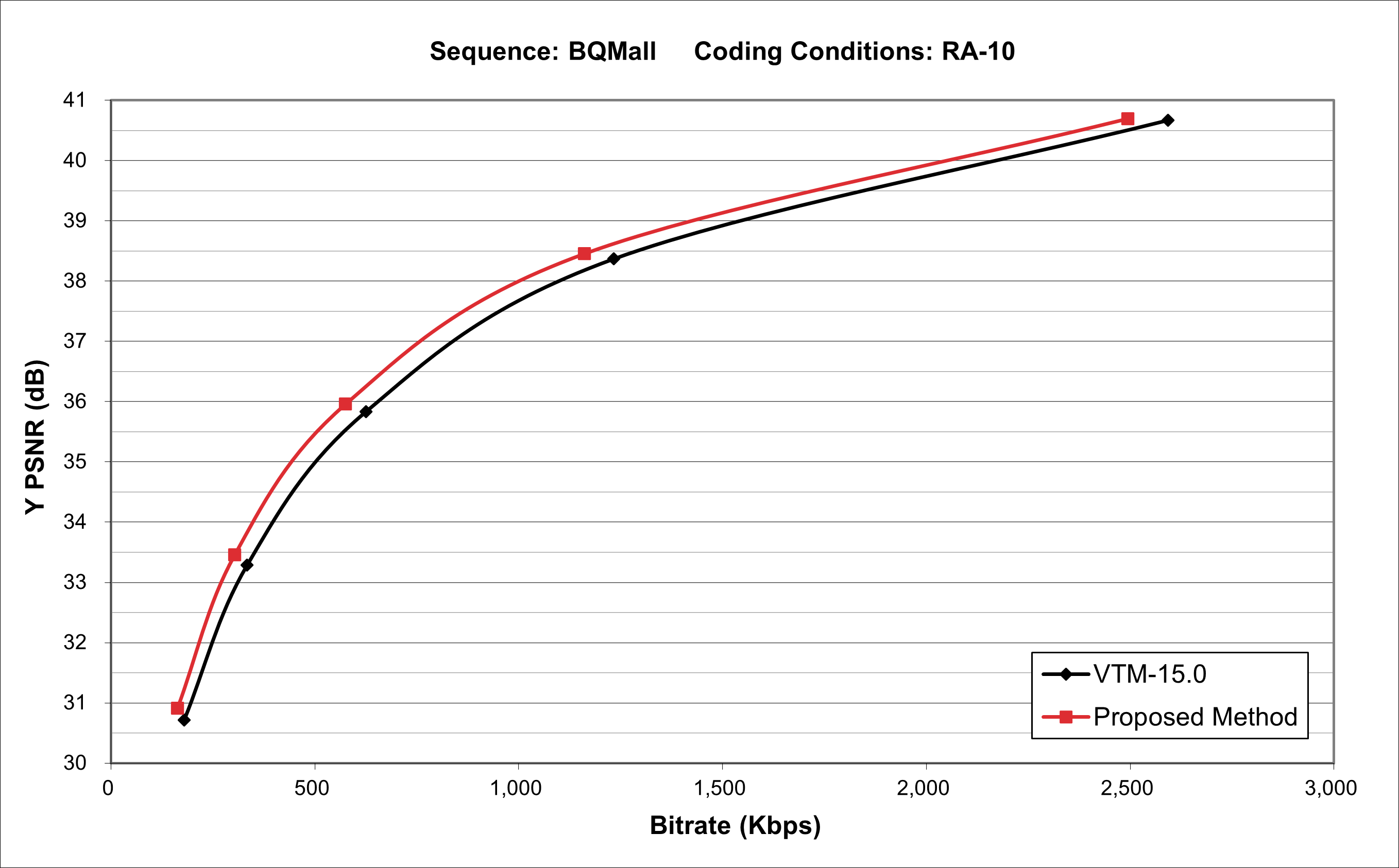}}
\quad
\subfigure[Class D - BasketballPass]{
\includegraphics[width=8.3cm]{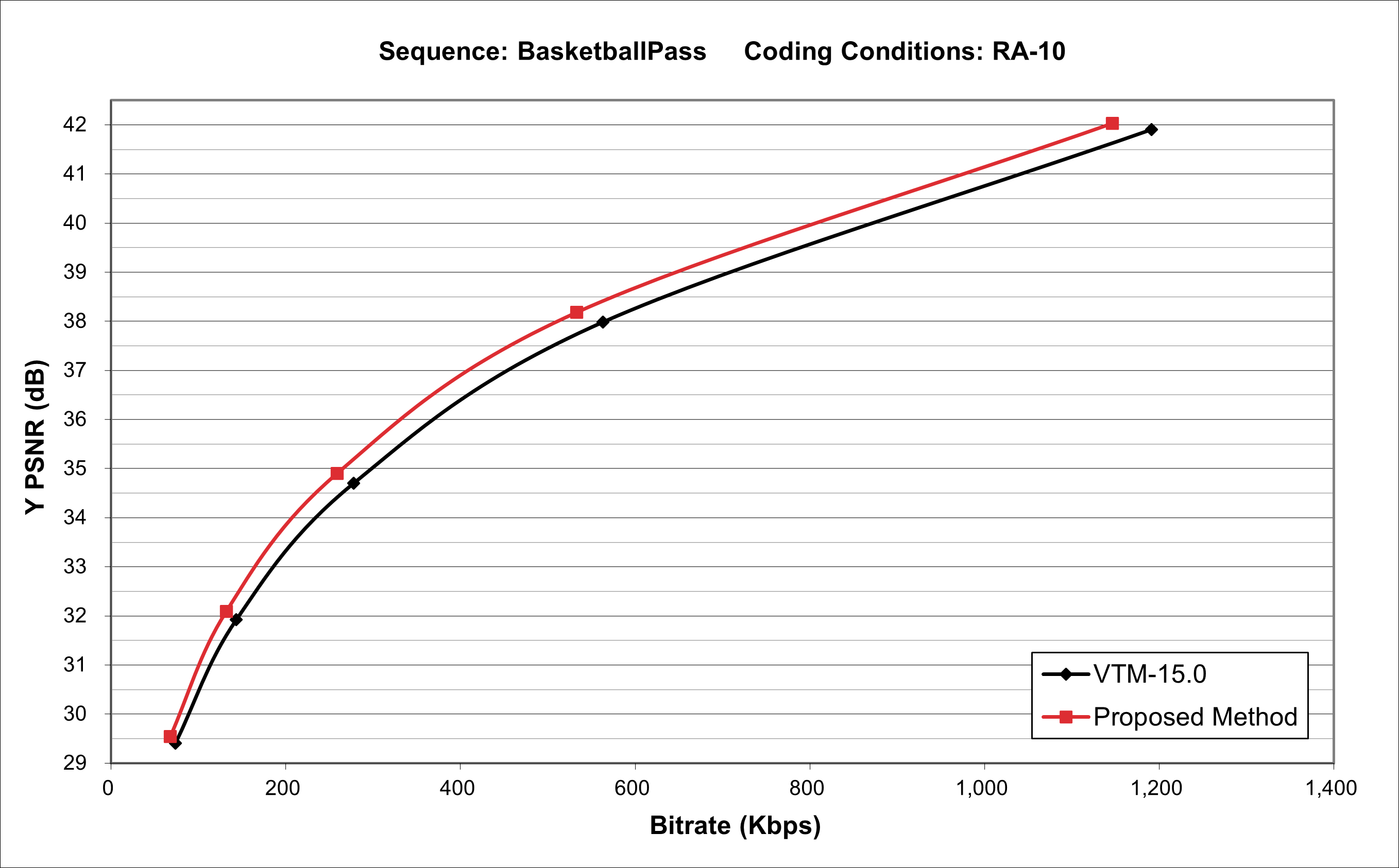}}
\end{center}
\caption{Four examples of RD curve on the sequences \emph{Tango2}, \emph{BasketballDrive}, \emph{BQMall}, and \emph{BasketbalPass}. All the sequences are encoded under RA configuration. Red curves denote the experimental results with the proposed method, while black curves represent the results on VTM-15.0.}
\label{RDCurve}
\end{figure*}

The Rate-Distortion (RD) curves for four distinct sequences at various resolutions are illustrated in Fig.\ref{RDCurve}. Each point on the curve depicts the distortion level introduced by the encoder to the coded video signal at a target bitrate. The red curves depict the results of our proposed method, while the black curves represent the results of VTM-15.0. As depicted in Fig.\ref{RDCurve}, the red points are located to the top left of the black ones, indicating that our approach achieves lower bitrates while having higher image quality. This perspective is proved by Table \ref{VTM-Bitrate-PSNR}, which highlights the average bitrate reduction and average Y-PSNR improvement achieved by the proposed method compared to VTM-15.0. Our method demonstrates a notable 4.08\% average bitrate reduction. Regarding video quality enhancement, Table \ref{VTM-Bitrate-PSNR} reveals an average Y-PSNR gain of approximately 0.29\%.

\begin{figure*}
\begin{center}
\subfigure[Class C - BQMall - QP27]{
\includegraphics[width=7.9cm]{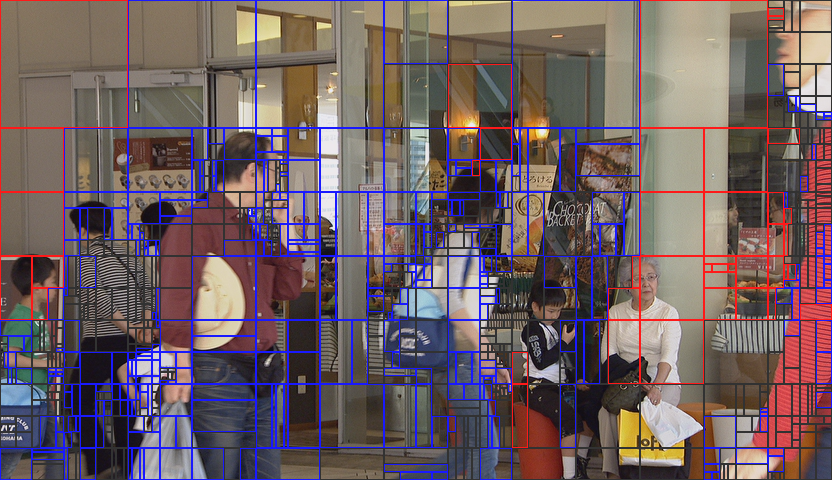}}
\quad
\subfigure[Class C - BQMall - QP37]{
\includegraphics[width=7.9cm]{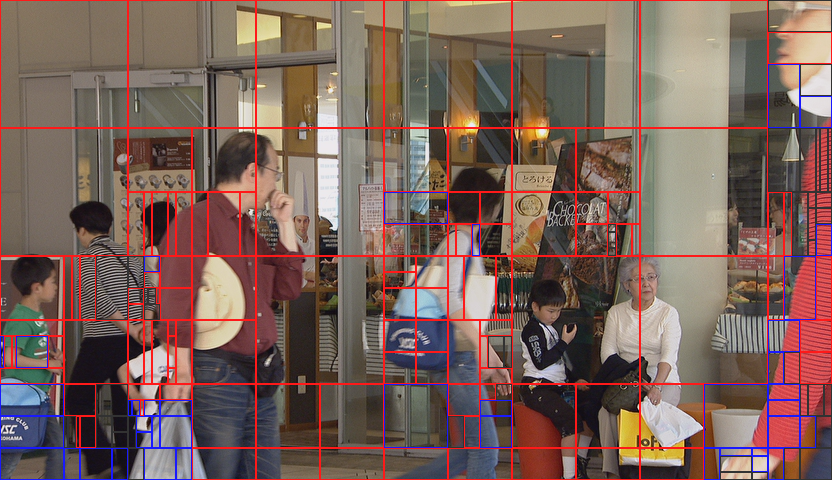}}
\quad
\subfigure[Class D - BasketballPass - QP27]{
\includegraphics[width=7.9cm]{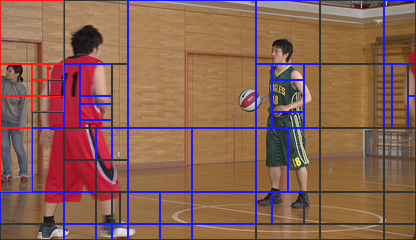}}
\quad
\subfigure[Class D - BasketballPass - QP37]{
\includegraphics[width=7.9cm]{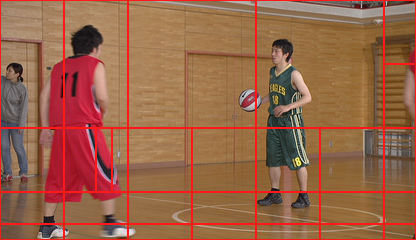}}
\end{center}
\caption{Four examples of reference frame selection. The red rectangle is the CU, whose two reference frames are the synthesized frames. The blue rectangle CU represents only one reference frame is the synthesized frame. The top two figures show the $26^{th}$ frame in \emph{BQMall} encoded with QP 27 and 37, while the bottom two figures are the $1^{st}$ frame in \emph{BasketballPass}.}
\label{CUs-Selection}
\end{figure*}

Each to-be-coded CU searches for matching reference blocks in two reference frames during bi-prediction. The more similarity between to-be-coded CUs and reference blocks, the fewer bitrates are required for encoding the residuals. Fig.\ref{CUs-Selection} illustrates the reference block selection for each CU under RA configuration. Rectangles of different colors represent different reference block usages of to-be-coded CUs. Red rectangles indicate that both reference blocks for the to-be-coded CU are from synthesized frames, blue rectangles signify that only one reference block is from a synthesized frame, and black rectangles denote that neither of the two reference blocks is synthesized. As shown in Fig.\ref{CUs-Selection}, many to-be-coded CUs choose synthesized blocks as references, especially at lower quality levels (higher QP). It is evident that synthesized frames provide good references for to-be-coded frames besides those original reconstructed frames.

The complexity analyses of our proposed method are provided in Table \ref{VTM-Time}. The time complexity is measured by evaluating the computational time ratio between our method and VTM-15.0. Compared to VTM-15.0, our method exhibits an average encoding complexity increase of 272\%. It should be noted that all times are evaluated using a CPU. The time increment primarily arises from the network inference of RFS, PFE, and JISE, as well as the rate-distortion optimization (RDO) processes for the synthesized reference frames. On the decoder side, only the network inference requires additional time. Regarding network complexity, the number of parameters is 9.8M, and the number of multiply-accumulates per pixel is 3792K.

\subsection{Ablation Study}

\begin{table*}
\caption{Tool-off Performance Comparison of the Proposed Method under RA Configuration}
\label{Ablation-Performance}
\setlength{\tabcolsep}{10pt}
\renewcommand\arraystretch{1.2}
\begin{center}
\begin{tabular}{c|ccccccccc}
\hline
\multirow{3}{*}{\textbf{Class}} & \multicolumn{9}{c}{\textbf{BD-rate}}                                                                                                                                                                                                                                     \\ \cline{2-10} 
                                & \multicolumn{3}{c|}{\textbf{(a) Without RFS}}                              & \multicolumn{3}{c|}{\textbf{(b) Without PFE}}                              & \multicolumn{3}{c}{\textbf{(c) Without Joint Training}}                  \\ \cline{2-10} 
                                & \textbf{Y-PSNR}      & \textbf{U-PSNR}     & \multicolumn{1}{c|}{\textbf{V-PSNR}}     & \textbf{Y-PSNR}      & \textbf{U-PSNR}     & \multicolumn{1}{c|}{\textbf{V-PSNR}}     & \textbf{Y-PSNR}      & \textbf{U-PSNR}     & \textbf{V-PSNR}     \\ \hline
\textbf{Class A1}               & 5.00\%          & 8.37\%                & \multicolumn{1}{c|}{8.89\%}                & 1.43\%          & 4.03\%                & \multicolumn{1}{c|}{4.88\%}                &  -0.08\%          & 1.16\%                & 0.11\%                \\
\textbf{Class A2}               & 6.77\%          & 11.60\%                & \multicolumn{1}{c|}{8.97\%}                & 0.96\%          & 3.97\%                & \multicolumn{1}{c|}{2.31\%}                & 0.72\%          & 1.31\%                & 1.41\%                \\
\textbf{Class B}                & 5.26\%          & 10.62\%          & \multicolumn{1}{c|}{9.62\%}          & 1.27\%          & 4.73\%          & \multicolumn{1}{c|}{3.21\%}          & 0.66\%          & 1.76\%          & 0.70\%          \\
\textbf{Class C}                & 7.26\%          & 15.23\%          & \multicolumn{1}{c|}{15.24\%}          & 1.36\%          & 6.53\%          & \multicolumn{1}{c|}{6.94\%}          & 0.45\%          & 1.93\%          & 1.56\%          \\
\textbf{Class D}                & 8.02\%          & 15.74\%          & \multicolumn{1}{c|}{16.67\%}          & 1.25\%          & 6.47\%          & \multicolumn{1}{c|}{6.78\%}          & 0.36\%          & 2.27\%          & 1.65\%          \\ \hline
\textbf{Average}                & \textbf{6.46\%} & \textbf{12.47\%} & \multicolumn{1}{c|}{\textbf{12.07\%}} & \textbf{1.26\%} & \textbf{5.24\%} & \multicolumn{1}{c|}{\textbf{4.87\%}} & \textbf{0.44\%} & \textbf{1.74\%} & \textbf{1.10\%} \\ \hline
\end{tabular}
\end{center}
\end{table*}

\begin{table}
\caption{Tool-off Runtime Comparison of the Proposed Method under RA Configuration}
\label{Ablation-Runtime}
\setlength{\tabcolsep}{5pt}
\renewcommand\arraystretch{1.2}
\begin{center}
\begin{tabular}{c|cccc}
\hline
\multirow{3}{*}{\textbf{Class}} & \multicolumn{4}{c}{\textbf{Runtime}}                                                                                                                                                                                                                                     \\ \cline{2-5} 
                                & \multicolumn{2}{c|}{\textbf{(a) Without RFS}}                              & \multicolumn{2}{c}{\textbf{(b) Without PFE}}                               \\ \cline{2-5} 
                                & \textbf{Encoding}      & \multicolumn{1}{c|}{\textbf{Decoding}}     & \textbf{Encoding}      & \textbf{Decoding}        
                                      \\ \hline
\textbf{Class A1}               & 61\%          & \multicolumn{1}{c|}{33\%}                & 91\%          & 85\%                               \\
\textbf{Class A2}               & 63\%          & \multicolumn{1}{c|}{33\%}                & 91\%          & 85\%                               \\
\textbf{Class B}                & 58\%          & \multicolumn{1}{c|}{32\%}                & 89\%          & 84\%                               \\
\textbf{Class C}                & 70\%          & \multicolumn{1}{c|}{33\%}                & 92\%          & 84\%                               \\
\textbf{Class D}                & 70\%          & \multicolumn{1}{c|}{33\%}                & 92\%          & 85\%                               \\ \hline
\textbf{Average}                & \textbf{66\%} & \multicolumn{1}{c|}{\textbf{33\%}}       & \textbf{91\%} & \textbf{85\%}                      \\ \hline
\end{tabular}
\end{center}
\end{table}

\begin{table*}
\caption{BD-rate Reduction Comparision between Existing Methods and the Proposed Method under RA Configuration}
\label{Comparision-RA}
\setlength{\tabcolsep}{4pt}
\renewcommand\arraystretch{1.2}
\begin{center}
\resizebox{0.9\textwidth}{!}{
\begin{tabular}{c|cccccccccccc}
\hline
\multirow{3}{*}{\textbf{Class}} & \multicolumn{12}{c}{\textbf{BD-rate}}                                                                                                                                                                                 \\ \cline{2-13} 
                                & \multicolumn{3}{c|}{\textbf{Hu \emph{et al.} \cite{Hu-ACM2023}}}                                      & \multicolumn{3}{c|}{\textbf{Jia \emph{et al.} \cite{Jia-TCSVT2023}}}           & \multicolumn{3}{c|}{\textbf{Ours (Without PFE) }}                           & \multicolumn{3}{c}{\textbf{Ours}}                    \\ \cline{2-13} 
                                & \textbf{Y-PSNR}       & \textbf{U-PSNR}       & \multicolumn{1}{c|}{\textbf{V-PSNR}}       & \textbf{Y-PSNR}       & \textbf{U-PSNR}       & \multicolumn{1}{c|}{\textbf{V-PSNR}}       & \textbf{Y-PSNR}       & \textbf{U-PSNR}       & \multicolumn{1}{c|}{\textbf{V-PSNR}}       & \textbf{Y-PSNR}       & \textbf{U-PSNR}        & \textbf{V-PSNR}        \\ \hline
Class A1                        & -0.74\%          & -1.85\%          & \multicolumn{1}{c|}{-2.35\%}          & -3.30\%          & -7.14\%          & \multicolumn{1}{c|}{-7.48\%} & -4.15\%          & -8.09\%          & \multicolumn{1}{c|}{-8.82\%}          & -5.22\%          & -10.81\%           & -12.14\%           \\
Class A2                        & -0.70\%          & -3.23\%          & \multicolumn{1}{c|}{-2.52\%}          & -4.92\%          & -10.73\%          & \multicolumn{1}{c|}{-8.66\%} & -6.20\%          & -11.40\%          & \multicolumn{1}{c|}{-9.37\%}          & -7.14\%          & -14.02\%          & -10.94\%           \\
Class B                         & -0.81\%          & -2.88\%          & \multicolumn{1}{c|}{-3.33\%}          & -3.50\%          & -8.97\%          & \multicolumn{1}{c|}{-8.70\%} & -4.56\%          & -10.80\%          & \multicolumn{1}{c|}{-9.91\%}          & -5.72\%          & -14.08\%           & -12.03\%           \\
Class C                         & -2.13\%          & -4.76\%          & \multicolumn{1}{c|}{-4.85\%}          & -4.92\%          & -11.75\%          & \multicolumn{1}{c|}{-12.02\%} & -6.08\%          & -13.32\%          & \multicolumn{1}{c|}{-13.44\%}          & -7.36\%          & -17.68\%          & -18.36\%          \\
Class D                         & -3.08\%          & -5.24\%          & \multicolumn{1}{c|}{-6.90\%}          & -6.84\%          & -12.69\%          & \multicolumn{1}{c|}{-14.12\%} & -7.34\%          & -13.75\%          & \multicolumn{1}{c|}{-15.20\%}          & -8.55\%          & -18.37\%          & -20.20\%          \\ \hline
\textbf{Average}                & \textbf{-1.54\%} & \textbf{-3.67\%} & \multicolumn{1}{c|}{\textbf{-4.12\%}} & \textbf{-4.69\%} & \textbf{-10.32\%} & \multicolumn{1}{c|}{\textbf{-10.34\%}}  & \textbf{-5.66\%} & \textbf{-11.62\%} & \multicolumn{1}{c|}{\textbf{-11.51\%}} & \textbf{-6.80\%} & \textbf{-15.22\%} & \textbf{-14.93\%} \\ \hline
\end{tabular}}
\end{center}
\end{table*}

The ablation study in this paper employs a ``tool-off'' test procedure. Specifically, the ``tool-off'' test for a given tool involves comparing the VTM software configured to disable that particular tool with the regular VTM software (where the tool would be enabled)\cite{Pfaff-TCSVT2021}. The experimental results are quantified using the PSNR-based BD-Rate. If a tool brings the compression improvement, disabling it will result in a positive BD-Rate value. In the ``tool-off'' test, we use the proposed method as the anchor and individually disable RFS, PFE, and the joint training strategy of RFS and PFE to examine their respective effects on the proposed method. The performance comparison of the ``tool-off'' test is illustrated in Table \ref{Ablation-Performance}, and the runtime comparison is shown in Table \ref{Ablation-Runtime}. 

\subsubsection{Reference Frame Synthesis}
RFS generates a virtual reference frame resembling the to-be-coded frame, effectively reducing residuals. We conduct a ``tool-off'' test to assess the significance of RFS, where RFS is disabled, and the proposed method serves as the anchor. The performance comparison, detailed in Table \ref{Ablation-Performance}(a), reveals performance losses of 6.46\%/12.47\%/12.07\% across three components when RFS is disabled. Simultaneously, as indicated in Table \ref{Ablation-Runtime}(a), encoding and decoding times are reduced to 66\% and 33\%, respectively. RFS is the primary driver of performance improvement in our methodology and constitutes a substantial portion of the total runtime.

\subsubsection{Post-processing Filter Enhancement}
PFE operates out of the encoding and decoding loops, impacting only the final reconstructed frames. PFE will alleviate artifacts introduced by block partitioning and quantization. To evaluate the impact of PFE, we turn it off during inference, with the proposed method serving as the anchor. As depicted in Table \ref{Ablation-Performance}(b), the absence of PFE leads to a noticeable decline in the performance of 1.26\%/5.24\%/4.87\%. The more significant performance drop in chroma components implies that PFE is particularly effective in mitigating artifacts within chroma components. As illustrated in Table \ref{Ablation-Runtime}(b), encoding time and decoding time are reduced to 91\% and 85\%, respectively. This indicates that PFE does not significantly contribute to the overall complexity of our approach.

\subsubsection{Joint Training Strategy}
As described in Section \ref{Training-Strategy}, the synthesis and enhancement pipelines are jointly trained within a single STENet, significantly simplifying the training cost. To investigate the impact of joint training on performance, we separately trained two STENets. For the first STENet, we only supervise the synthesized frame (second frame) and maintain all training settings unchanged. For the second STENet, we supervise the enhanced frames (first and third frames) and keep the original training settings.

Regarding network inference, the first STENet's synthesis pipeline is utilized in RFS, while the second STENet's enhancement pipeline is utilized in PFE. As for JISE, the single STENet is replaced by a cascade of two pipelines. As shown in Table \ref{Ablation-Performance}, training two separate STENets for RFS and PFE resulted in a performance decrease of 0.44\%/1.74\%/1.10\%. It indicates that the joint training strategy not only simplifies training costs but also enhances performance in the proposed method.

\subsection{Comparison with Existing Methods}
We compare the proposed method with two other RFS approaches \cite{Hu-ACM2023} and \cite{Jia-TCSVT2023}. Hu et al. \cite{Hu-ACM2023} introduce the Error-Corrected Auto-Corrected Network (ECAR-Net) for inter prediction in video coding. Jia et al. \cite{Jia-TCSVT2023} present a deep reference frame generation method, incorporating an interpolation network to enhance bi-directional prediction. Both methods are integrated into VTM-15.0 and operated under RA configuration. Experimental results are detailed in Table \ref{Comparision-RA}, with results for \cite{Hu-ACM2023} and \cite{Jia-TCSVT2023} obtained from their respective papers. We maintain the same QP values to align their experimental results as \cite{Hu-ACM2023} only provides QP values for 22, 27, 32, and 37. For a fair comparison, we report the performance of the proposed method with PFE disabled. Evidently, with PFE disabled, our method outperforms theirs in each class. When PFE is enabled, our method exhibits even more significant performance gains. On average, a coding efficiency enhancement of up to 6.80\% is achieved in the Y component, surpassing the 1.54\% and 4.69\% improvements reported in \cite{Hu-ACM2023} and \cite{Jia-TCSVT2023}, respectively. The proposed method's performance in the U and V components far exceeds that of \cite{Hu-ACM2023} and \cite{Jia-TCSVT2023}.

\section{Conclusions}
This paper presents joint RFS and PFE for VVC. We propose a well-designed STENet, in which two input frames are enhanced through its enhancement pipeline and utilized to synthesize an intermediate frame through its synthesis pipeline. During RFS, two reconstructed frames are sent into STENet's synthesis pipeline to synthesize a virtual reference frame, which is inserted into two RPLs as additional reference frames. During PFE, two reconstructed frames are sent into STENet's enhancement pipeline to alleviate their artifacts and distortions. In addition, we propose JISE to reduce inference time. Meanwhile, the proposed method only needs one joint training, which saves lots of resources. Integrated into the VVC reference software VTM-15.0, the proposed method could achieve -7.34\%/-17.21\%/-16.65\% PSNR-based BD-rate on average for three components under RA configuration. Experimental results demonstrate the effectiveness and superiority of our proposed method over existing methods.
\bibliographystyle{ieeetr}
\bibliography{ref}

\end{document}